# Beyond Accessibility: How Intelligent Assistive Technologies Improve Activities of Daily Life for Visually Impaired People in South Africa

Ronaldo Nombakuse
University of Cape Town
nmbxol002@uct.ac.za

Nils Messerschmidt
SnT, University of Luxembourg
nils.messerschmidt@uni.lu

Pitso Tsibolane
University of Cape Town
pitso.tsibolane@uct.ac.za

Muhammad Irfan Khalid
University of Agder
muhammad.i.khalid@uia.no

## Abstract

*Our study explores how intelligent assistive technologies (IATs) can enable visually impaired people (VIPs) to overcome barriers to inclusion in a digital society to ultimately improve their quality of life. Drawing on the Social Model of Disability (SMD), which frames disability as a consequence of social and institutional barriers rather than individual impairments, we employ semi-structured interviews and an online qualitative survey with n=61 VIPs in South Africa. Using descriptive statistics and Qualitative Comparative Analysis (QCA), we uncover nine configurations, clustered along three broader combinations of conditions, that support and hinder IAT-mediated inclusion. Most notably, we identify that autonomy of VIPs and accessibility of IATs are primary predictors of IAT's ability to achieve social participation. Our findings contribute to Information Systems (IS) literature at the intersection of technology and social participation. We further formulate implications for research and policymakers to foster social inclusion of VIPs in the Global South.*

## 1. Introduction

Digital societies present significant challenges for individuals with visual impairments, often leading to persistent social exclusion (Oliver, 1983, 2013). Despite technological advancements, many online products and services remain inaccessible, which creates an imbalance where powerful assistive technology is available, but the fundamental digital infrastructure still poses significant barriers (Kugler, 2020; Balakrishnan, 2022). This exclusion is not inherent to an individual's impairment but is caused by societal barriers, a lack of accessible services, and environments that do not accommodate people with disabilities, as highlighted by the social model of disability (Joshi & Pappageorge, 2023). This is evident in various domains, including *navigation in urban environments* (Rodriguez-Sanchez & Martinez-Romo, 2017; Balakrishnan, 2022), *access to public information* (Rodriguez-Sanchez & Martinez-Romo, 2017), *transport systems* (Muñoz et al., 2016), and *participation in online activities and education* (Joshi & Pappageorge, 2023; Senjam, 2021). Addressing these barriers is crucial for ensuring equal access and participation (Joshi & Pappageorge, 2023).

Intelligent assistive technologies (IATs) can offer substantial potential to mitigate these challenges by facilitating independent living and empowering autonomy for millions of blind and visually impaired individuals (Rodriguez-Sanchez & Martinez-Romo, 2017; Muñoz et al., 2016; Balakrishnan, 2022). They serve as a key platform for accessible applications and features (Rodriguez-Sanchez & Martinez-Romo, 2017; Senjam, 2021; Balakrishnan, 2022) that enable users to perform daily activities such as making telephone calls, sending text messages, browsing the internet, and accessing social media (Rodriguez-Sanchez & Martinez-Romo, 2017; Senjam, 2021).

At least 2.2 billion people worldwide have near or distance vision impairment (WHO, 2023). Africa bears a substantial burden of visual impairment, accounting for an estimated 15.3% of the global blind population. This represents approximately 26.3 million individuals across the continent, comprising 5.9 million people with blindness and 20.4 million with low vision conditions (WHO, 2024).

The prevalence of visual impairment in Africa highlights the significant public health challenge facing the region, where nearly one in six of the world's blind individuals reside (WHO, 2024). This demographic distribution highlights the disproportionate impact of vision-related disabilities on African populations and emphasizes the critical need for comprehensive eye care services and preventive interventions across the continent (Addo et al., 2021). In low- and middle-income (LMI) contexts, notably South Africa, disability and poverty exhibit a bidirectional relationship, creating a cyclical pattern of compounding deprivation (Tsibolane & Nombakuse, 2024). Individuals with visual impairment are particularly susceptible to this phenomenon. Among South Africa's estimated 724,000 persons with visual impairment, unemployment rates reach approximately 97%, demonstrating the significant economic marginalization experienced by this population.

This high unemployment rate illustrates the systemic barriers that prevent individuals with visual impairment from accessing economic opportunities, thereby perpetuating cycles of poverty and social exclusion. The intersection of visual disability and economic disadvantage reflects broader structural inequalities, whereby inadequate support systems, limited accessibility infrastructure, and discriminatory practices converge to restrict meaningful participation in the formal economy for persons with disabilities.

Consequently, the focus on South Africa is justified by significant challenges that are relevant and timely. Although research has explored barriers faced by individuals with disabilities in a South African educational institution, such as inaccessibility to facilities and exclusion from academic activities, and made recommendations for addressing these (Joshi & Pappageorge, 2023), a comprehensive understanding of how IATs are currently shaping their pathways to an improved quality of life across various domains is needed. Therefore, this study seeks to address this gap by investigating: *How do intelligent assistive technologies enable activities of daily living for visually impaired people in South Africa?*

In particular, the study seeks to explain the factors that enable the activities of daily living for visually impaired people. This paper positions itself within the broader Information Systems (IS) research community that focuses on social inclusion, accessibility, and digital participation. By adopting a user-focused approach (Rodriguez-Sanchez & Martinez-Romo, 2017) and examining how technology interacts with societal structures (i.e., reflecting the social model) (Joshi & Pappageorge, 2023), this work contributes to understanding how to leverage digital solutions to overcome barriers and promote equality for individuals with disabilities (Rodriguez-Sanchez & Martinez-Romo, 2017; Kim et al., 2016; Kugler, 2020; Joshi & Pappageorge, 2023; Balakrishnan, 2022). While considerable research exists on specific assistive devices and accessibility features (Rodriguez-Sanchez & Martinez-Romo, 2017; Kim et al., 2016), a need for studies that explore the pathways through which the use of these technologies translates into tangible improvements in the quality of life for visually impaired people remains (Joshi & Pappageorge, 2023). Hence, with the use of Qualitative Comparative Analysis (QCA), we uncover nine configurations, clustered along three broader combinations of conditions, that support and hinder IAT-mediated inclusion. Most notably, we identify that autonomy of VIPs and accessibility of IATs are primary predictors of IATs' ability to achieve social participation. Highlighting the significance of personalized support and local communities, we formulate implications for future research and policymakers to achieve social inclusion of VIPs.

## 2. Theoretical Background

### 2.1. Social Model of Disability

The Social Model of Disability (SMD) (Oliver, 1983, 2013) challenges the traditional (i.e., functional) medical model (Olkin, 2002; Xie et al., 2020; Haegele & Hodge, 2016), which traditionally views disability as an inherent problem in the individual stemming from biological defects, limitations, or impairments that need to be cured, altered, or normalized (Cao et al., 2020; Vimalan et al., 2024). Figure 1 depicts this differentiation.

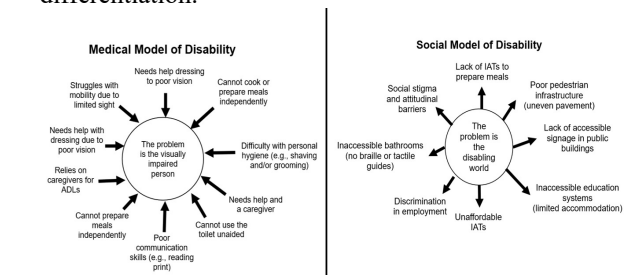

**Figure 1. The Medical vs. Social Model of Disability**

The SMD posits that disability is caused by how society is organized, rather than by a person's impairment (Cao et al., 2020). According to Vimalan et al. (2024), the central idea is the distinction between *disability* and *impairment*. Impairment refers to the physical, sensory, or cognitive characteristics of an individual, while disability is the disadvantage or restriction of activity caused by societal barriers and the lack of accommodation for people with impairments. The SMD further argues that society disables impaired people by isolating them, making disability a societal construct (Sanderson et al., 2024).

Thus, the problem lies not in the individual's condition but in the environmental, social, cultural,

and political structures that limit participation and fail to accommodate differences (Xie et al., 2020). From the SMD perspective, accommodations are seen as actions to remove barriers that restrict choices through a collaborative process (Cao et al., 2020). The social model emphasizes the right of people with impairments to participate in all aspects of life fully and calls for designing inclusive social structures and environments (Cao et al., 2020).

Technology scholars have increasingly drawn on the SMD to understand and address the challenges faced by people with disabilities in interacting with technology and digital environments (Cao et al., 2020; Fota & Schramm-Klein, 2024; Sanderson et al., 2024; Vimalan, Zimmer, & Drews, 2024; Xie et al., 2020). The SMD has been extensively applied across various technology research domains to reframe disability as a societal issue rather than an individual limitation. This perspective has proven valuable in multiple contexts, demonstrating its flexibility and importance in addressing systemic barriers.

In digital library environments, the SMD has been used to frame the vulnerabilities of VIPs, positioning access limitations as social barriers created by improper design rather than inherent consequences of visual impairment (Xie et al., 2020). Sight-centered designs, complex structures, and multimedia formats in existing digital libraries contradict non-visual access modes, highlighting key barriers from a social model perspective (Xie et al., 2020).

The model has informed IS design for workplace accommodation processes, recognizing that traditional systems often fail to address social challenges like conflicting stakeholder interests (Cao et al., 2020). This necessitates designing information systems from a social model perspective, applying a critical social inclusion lens to system design (Cao et al., 2020).

Inclusive IS design has also been analyzed through social model principles to demonstrate that existing approaches like Universal Design predominantly reflect the functional (medical) model (Vimalan, Zimmer, & Drews, 2024). The social model perspective highlights overlooked social and structural factors which typically leads to conceptualisation of social inclusion principles that promote diversity and encourage participation of impaired individuals in design processes (Vimalan, Zimmer, & Drews, 2024).

Digital inclusion research has been critiqued using SMD-related models to advocate for a shift from deficit (functional) approaches toward strengths-based digital justice perspectives that remove systemic barriers to enable equity and empowerment (Sanderson et al., 2024). Additionally, the model has been applied to study smart technologies like digital voice assistants, examining how they enable social participation by removing environmental barriers (Fota & Schramm-Klein, 2024).

When applied to studies on intelligent assistive technologies for visually impaired persons, the social model provides a framework for understanding how technology interventions address disability as societal restriction rather than attempting to "fix" visual impairment (Cao et al., 2020; Fota & Schramm-Klein, 2024; Vimalan, Zimmer, & Drews, 2024; Xie et al., 2020). Barriers faced by VIPs include inaccessible resources such as websites, sight-centered interfaces, and a lack of non-visual alternatives, are understood as environmental, and societal barriers created by human choices rather than inherent consequences of visual impairment (Xie et al., 2020).

This study views IATs through the SMD lens to demonstrate how these technologies empower VIPs by addressing systemic barriers in the digital and social environment, enabling their activities of daily living (ADLs) and improving their quality of life.

## 2.2. Activities of Daily Living

The concept of activities of daily living (ADLs) was initially developed by Katz (1983) to evaluate whether an individual has the capabilities to perform fundamental tasks independently (Lo et al., 2024). Initially, ADLs were utilized in public health to evaluate the functional status of the elderly and people living with disabilities (PWDs) (Zhu et al., 2020). ADLs include fundamental self-care skills such as eating, bathing, dressing, toileting, and mobility, which are the cornerstone of independent living (Edemekong et al., 2023). ADLs serve as critical indicators of functional status and quality of life. With particular significance for visually impaired persons (VIPs) who face unique structural barriers in performing these essential routines (Mercan et al., 2021). For VIPs, the inability to independently complete ADLs often results in increased dependence on caregivers (Gao et al., 2024), heightened safety risks (McGrath et al., 2025), and diminished overall well-being (Datt et al., 2017). Healthcare professionals worldwide recognize ADL assessment as a vital tool for predicting healthcare needs, including nursing home admissions, hospitalization requirements, and home care services (Khan & Khusro, 2021). The progressive nature of vision loss, whether through aging or chronic conditions, can systematically erode one's capacity to maintain independence in daily activities, creating a cascade of challenges that extend beyond the physical realm to encompass psychological and social dimensions of health.

There has been a plethora of IS studies (Hubner et al., 2022; Shethia et al., 2023; Zhu et al., 2020) that use ADLs to explain how vulnerable people achieve their functional status (Aminparvin et al., 2025). The five basic ADLs are as follows.

- *Transferring*: The ability to move from one place to another independently of their caregiver,
- *Dressing*: The ability to select suitable outfits and dress oneself,
- *Feeding*: The ability to choose the desired food and feed oneself,
- *Toileting*: The ability to visit, utilize, clean, and return from the bathroom, and
- *Bathing*: The ability to maintain personal hygiene, viz., bathing, brushing teeth, grooming, manicuring, and pedicuring.

**2.3. Intelligent Assistive Technologies**

IATs present opportunities to bridge gaps and restore autonomy for VIPs, as these technologies are specifically designed to improve the quality of life for individuals with visual impairments (Edemekong et al., 2023). Through IATs such as smart navigation systems, voice-activated interfaces, tactile feedback devices, and AI-powered voice and audio recognition tools (Gao et al., 2024; McGrath et al., 2025; Shethia et al., 2023; Zhu et al., 2020). VIPs can overcome traditional barriers that have historically limited their independence in performing the five basic ADLs: washing oneself, using the toilet independently, dressing oneself, preparing and feeding oneself, and mobility (Pizarro-Pennarolli et al., 2021). These technologies not only enhance safety and efficiency in performing ADLs but also promote dignity and self-determination by reducing reliance on human assistance (Lo et al., 2024). Integrating IATs into daily routines represents empowerment from accommodation-based approaches to empowerment-focused solutions (McGrath et al., 2025). Enabling VIPs to maintain their preferred living arrangements (Tsibolane & Nombakuse, 2024), participate more fully in community life (Datt et al., 2017), and achieve optimal quality of life outcomes while preserving their autonomy and personal agency (Hubner et al., 2022; Shethia et al., 2023).

There are various IATs that VIPs use on their ADLs. Table 1 presents 15 IATs that support VIPs across five ADL categories. *Mobility and transferring* are supported by 11 IATs, featuring comprehensive visual recognition tools (e.g., Be My Eyes, TapTap See, InVision AI) and specialized devices (e.g., Smart Cane). *Feeding* activities utilize 2 IATs, notably Be My Eyes for meal preparation and Liquid Level Indicator for measuring liquids. *Dressing* is addressed by 6 technologies, primarily through multi-functional platforms. *Bathing and/or toileting* are supported by 2 IATs. These IATs demonstrate varied approaches from specialized single-function devices to comprehensive multi-ADL platforms (Datt et al., 2017; Hubner et al., 2022; Shethia et al., 2023), collectively enabling VIPs to maintain independence across essential activities of daily living (Aminparvin et al., 2025; McGrath et al., 2025).

**Table 1. Overview of primary IATs by ADL**

| ADLs | Primary IATs |
|---|---|
| Transferring | iOS Voiceover, Android Talkback, BlindSquare, Be My Eyes, Loadstone GPS, Lazerillo, Envision AI, TapTap See, Smart Cane, Invoice Connect, Invision AI |
| Dressing | Be My Eyes, Color Detector App, iOS Voiceover, Android Talkback, Needle Witch, Braille Watch |
| Feeding | Liquid Level Indicator, Be My Eyes |
| Bathing | Android Talkback, iOS Voiceover |
| Toileting | Android Talkback, iOS Voiceover |

The existing literature primarily focused on understanding the relationship between the usefulness of technologies for improving the activities of VIPs (Aminparvin et al., 2025; Datt et al., 2017; Hubner et al., 2022), and there have been consistent debates. Researchers have utilized several methods and tools to understand this relationship. A recent study by Fota & Schramm-Klein (2024) demonstrates the impact of IATs on people living with disabilities. This study uses qualitative content analysis to determine the barriers and drivers of utilization. Their study operates within a linear, categorical interpretive framework that treats influencing factors as independent, isolated themes rather than exploring how multiple conditions interact to produce adoption or rejection outcomes. It does not identify combinations of conditions (e.g., financial support, usability, independence) that lead to adoption (Fota & Schramm-Klein, 2024). Following the same lines, Vieira et al. (2022) explored the impact of voice assistant technologies on people with physical and visual impairments using thematic analysis to extract key themes from interviews, observations, and device logs. Their study follows a thematic, narrative framework that treats influencing factors such as independence, convenience, and inclusiveness as separate insights. However, it does not examine how combinations of conditions interact to produce outcomes that improve well-being. The use of QCA could have revealed causal configurations (e.g., impairment type, internet quality, family support) that are necessary or sufficient for enhanced well-being, offering a more configurational understanding of user experiences (Vieira et al., 2022). These studies reflect the potential of configurational analysis on these kinds of datasets to draw various conclusions.

The integration of technologies like computer vision and cloud computing has enabled fascinating innovations to resolve everyday barriers (Rodriguez-Sanchez & Martinez-Romo, 2017; Kugler, 2020; Balakrishnan, 2022). These include mobility applications designed for universal accessibility in both indoor and outdoor environments (Rodriguez-Sanchez & Martinez-Romo, 2017; Balakrishnan, 2022), AI-enhanced tools such as money readers (Kugler, 2020; Balakrishnan, 2022), screen reading software (Rodriguez-Sanchez & Martinez-Romo, 2017), and innovative interfaces using haptic feedback (vibrotactile) and gestures for interacting with large displays (Kim, Ren, Choi, & Tan, 2016). Furthermore, devices like smart canes (Balakrishnan, 2022), systems for independent public transport use (e.g., the SIMON project for multimodal navigation) (Muñoz et al., 2016), and refreshable Braille displays (Balakrishnan, 2022) demonstrate the many ways technology can support mobility, access to digital data, and education. These technologies are designed to increase independence, autonomy, and improve access to information (Rodriguez-Sanchez & Martinez-Romo, 2017; Kim et al., 2016; Muñoz et al., 2016; Balakrishnan, 2022; Senjam, 2021), thereby reducing dependency on caregivers (Aminparvin et al., 2025; Datt et al., 2017; Hubner et al., 2022; McGrath et al., 2025; Shethia et al., 2023).

## 3. Research Methodology

### 3.1. Data Collection

We applied a purposive and snowball sampling technique to recruit VIPs through community-based organizations and social media. We only recruited VIPs that currently utilise some type of an IAT, listed in Table 1. The data was collected using a semi-structured interview (Phase 1), and an online qualitative survey was administered (Phase 2). In the first phase, twenty-four (n=24) VIPs participated, as presented in Table 2. Before each interview, consent to record was requested from all VIPs. The duration for each interview was thirty minutes to an hour, and an iOS mobile phone was used for recording. All VIPs were allowed to provide their answers in their native languages, and all non-English interviews were manually translated into English. The audio data and interview transcripts (four to seven pages long) were securely stored on a cloud-based platform accessible only to the researchers. In the second phase, we utilized an online qualitative survey to support VIPs reluctant to be interviewed face-to-face and/or remotely. The online qualitative survey was designed on Qualtrics, given that the platform has text-to-speech (TTS) and screen magnification functionalities. All respondents were allowed to request assistance from the researchers if they experienced any challenges completing the survey. The respondents were also informed that they can complete the survey in their native languages. At least forty-nine VIPs completed the survey, and all partially completed surveys (N = 12) were removed from the data. The remaining surveys (N = 37) were downloaded and securely stored in the cloud. Both instruments were piloted with at least four VIPs (two low-visioned and two totally blind) to make amendments in preparation for the larger study.

**Table 2. Demographic information**

| Demographic Variable | Category | Number (N=61) |
|---|---|---|
| Age | 21-44 | 50 |
| | 45-65 | 11 |
| Education | High school | 33 |
| | Tertiary | 28 |
| Employment | Unemployed | 37 |
| | Employed | 24 |
| Vision Classification | Low vision | 15 |
| | Totally blind | 46 |
| Disability Grant | Yes | 53 |
| | No | 8 |

Table 2 presents the demographic information of 61 VIPs who participated in this study via semi-structured interviews and online qualitative surveys. At least 82% of VIPs who participated in the study were between the ages of 21 and 44, while only 18% were above the age of 45. Given the challenges faced by VIPs in South Africa, only 46% of VIPs had a tertiary qualification, which includes a national vocational certificate (NCV), diploma, and bachelor's degrees. Despite many possessing these qualifications, 61% were unemployed, which shows limited employment opportunities within this population. More than 75% of VIPs are within the classification of total blindness or no light perception (NLP), while 25% were low-visioned people. Surprisingly, a VIP can be employed and still be a recipient of disability grant, thus 87% indicated that they receive it.

### 3.2. Measures

For the measures of the QCA study, we make use of three broader types of constructs, these are ADLs,

IATs, and demographics that together explain the perceived usefulness of IATs (Figure 2). Table 3 provides the sources for each construct. There are five basic ADLs that we utilize as measures (Edemekong et al., 2023), to understand the IATs that VIPs use. In this study, paid apps refer to IATs utilized by VIPs through their smartphones, tablets, laptops, and desktops. Typically, VIPs pay a recurring subscription fee to use these apps (Shethia et al., 2023). For instance, in the text-to-speech context, while free apps such as NVDA or Android Talkback are available, Jaws is a paid app that is often perceived as superior.

Table 3. Overview of Constructs

| Construct | Source |
|---|---|
| **Activities of Daily Living** | |
| Transferring | (Aminparvin et al., 2025; Edemekong et al., 2023; Hubner et al., 2022; Khan & Khusro, 2021; Lo et al., 2024; Zhu et al., 2020) |
| Dressing | |
| Feeding | |
| Toileting | |
| Bathing | |
| **Technology Usage** | |
| Paid App Usage | (Balakrishnan, 2022; Datt et al., 2017; Gao et al., 2024; Shethia et al., 2023) |
| **Demographics** | |
| Education | (Addo et al., 2021 ; Joshi & Pappageorge, 2023; Tsibolane & Nombakuse, 2024) |
| Disability Grant | |

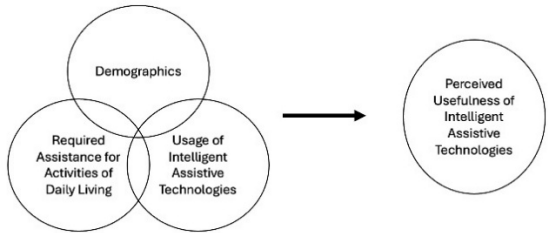

**Figure 2. Venn Diagram explaining Usefulness of IATs**

We also investigate demographic measures, namely, education and disability grants. Despite VIPs being qualified (Joshi & Pappageorge, 2023), they still cannot compete in the job market (Tsibolane & Nombakuse, 2024). Many rely solely on their monthly disability grant of R2 301 (Addo et al., 2021), which is not enough to afford IATs given their costs (Tsibolane & Nombakuse, 2024).

### 3.3. Qualitative Comparative Analysis

We applied QCA following the appropriate guidelines and recommendations by Mattke et al. (2022). QCA is a method that combines quantitative and qualitative analysis and aims to identify the conditions for an outcome to occur (Ragin, 2009). This method is appropriate for our data as it can also be used on small samples and allows us to investigate structurally different constructs (Pappas & Woodside, 2021). We used the fuzzy-set qualitative comparative analysis (fsQCA) software (Ragin & Davey, 2022) to compute the results. The method draws on the principle of equifinality and allows us to identify multiple solutions that explain the same outcome (Pappas & Woodside, 2021). Further, QCA draws on the principle of causal asymmetry, meaning that the conditions that explain the outcome are not the exact opposites of those that explain the absence of the outcome (Ragin, 2009). This enables us to infer an enhanced understanding of causal conditions and configurational relationships to identify the conditions contributing to the perceived usefulness of IATs, or the absence thereof, that have been used to assist VIPs.

Due to the inherent nature of our data, we used multi-value qualitative comparative analysis (mvQCA), a variant of QCA. MvQCA is superior to crisp-set qualitative comparative analysis (csQCA), as it allows for more than two values in conditions, thus reducing the likelihood of contradictory configurations (Vink & van Vliet, 2009). On the other hand, it can capture the specific causal role of intermediate categories in multi-value conditions, which fsQCA may struggle with (Haesebrouck, 2015).

## 4. Results

### 4.1. Measurements

To analyze the configurations leading to perceived usefulness of assistive technologies, several crisp-set conditions are used. In accordance with Figure 1, these are related to the ADLs, usage of assistive technologies, and demographic conditions. Table 4 presents those measures, as well as the meaning and the distribution within the dataset of the absence (0) and presence (1) for each condition. For ADLs, we find that those VIPs requiring assistance for feeding also require assistance for toileting and bathing, and vice versa. We were therefore able to merge those three ADLs into one condition.

To determine the value of the outcome variable, the following question was presented to participants in both the online qualitative survey and the interview: "*How useful do you think are technology-based devices on your daily activities as visually impaired person in South Africa?*"

For data calibration of qualitative data, a four-point scale is a common practice (Mattke et al., 2022). The participants in our dataset are distributed as follows: 9.8% at 0 (not useful at all); 24.6% at 0.33 (not very useful); 26.3% at 0.67 (somewhat useful); and 39.3% at 1 (very useful). Four researchers independently assessed the perceived usefulness of IATs based on qualitative responses. In case of disagreement, the value was assigned according to the average value closest to one of the four points.

Table 4. Overview of Measures

| Condition | Absence (0) | Presence (1) |
|---|---|---|
| **Activities of Daily Living** | | |
| Transferring (a) | No assistance needed (a) *0.689* (b) *0.557* (c) *0.672* | Assistance needed (a) *0.311* (b) *0.443* (c) *0.328* |
| Dressing (b) | | |
| Feeding / Toileting / Bathing (c) | | |
| **Technology Usage** | | |
| Paid App Usage | No access *(0.672)* | Access *(0.328)* |
| **Demographics** | | |
| Education | High school *(0.541)* | Tertiary *(0.459)* |
| Disability Grant | No *(0.131)* | Yes *(0.869)* |

After data calibration, several steps are to be taken before obtaining the QCA findings. The first step is to compute the truth table. The truth table has $2^k$ possible combinations, where $k$ equals the number of conditions used (Ragin, 2009). The six conditions in our analysis yield 64 possible combinations. We find 28 unique combinations in our dataset.

For datasets with smaller *n*, e.g., less than 150 observations, a frequency threshold of at least 2 is recommended (Fiss, 2011). Given the size and characteristics of our dataset, we determine 2 to be the most appropriate threshold. 16 combinations in our dataset fulfill the frequency threshold. The consistency threshold, which captures the degree to which a combination of conditions is consistently matched to an outcome (Fiss, 2011), is set to 0.75, slightly above the recommended 0.7 (Pappas & Woodside, 2021).

The Proportional Reduction in Inconsistency (PRI) consistency threshold, which captures how consistently a configuration leads to a specific outcome, is set to 0.5 (Pappas & Woodside, 2021).

Finally, the fsQCA software computes three different solutions, namely, complex, parsimonious, and intermediate. We use the intermediate solution to report the findings as it balances theoretical complexity and empirical knowledge (Ragin, 2009).

However, to enable us to draw more detailed insights from our analysis, we further use the parsimonious solution to determine the presence of core and peripheral conditions (Fiss, 2011).

### 4.2 mvQCA Findings

Table 5 presents the findings of the mvQCA analysis. We find a total of nine configurations. Five lead to positive perceived usefulness (1-5), with the remaining four configurations lead to negative perceived usefulness (6-9). The nine configurations can be grouped into three groups; these are as follows.

**1-3: High independence.** Configurations 1 to 3 capture individuals who perceive the IAT usage as useful, while retaining a high level of independence even without the use of technology. This can be observed by the absence of needs for assistance during travelling (e.g., individuals leverage public transport/ride-hailing services), as well as the absence of assistance required for feeding, bathing, and toileting. In each category, the absence of at least one of those ADLs is a core condition. Therefore, these individuals have lower requirements for IAT devices and services to make use of them. While in all configurations, individuals receive disability grants, it only constitutes a core condition in one configuration. Despite the grants, access to paid apps was either absent or not relevant. While the prevalence of high levels of education is core to one configuration, the absence of high levels of education is core to another. Therefore, education does not appear to affect the perception of this group.

**4-5: Low independence, high IAT accessibility.** Compared to the previous group, individuals falling within configurations 4 and 5 are found to be less independent, highlighted by the need for assisted travel (configuration 5) and assistance for feeding, bathing, and toileting, and/or dressing needs (configuration 4). However, none of these ADLs are core to those configurations. Instead, it was found that the core condition for those individuals is access to paid apps. Neither education nor disability grants were present in both configurations, highlighting the need for the family or community to bear the costs associated with access to paid apps. This shows that, while the group has higher requirements for IAT devices and services, they can meet those requirements by relying on high-end solutions, leading to a positive perception of IAT usage.

**6-9: Low independence, low IAT accessibility.** Configurations 6 to 9 are found to have negative perceived usefulness of IATs. Similarly to the previous group, these individuals have a high need for assistance across ADLs. Most prominently, the need for assisted travel is a core condition in three of the four configurations, while for the fourth one, feeding, bathing and toileting assistance is a core condition.

Table 5. Overview of mvQCA findings

| | Positive Perceived Usefulness | | | | | Negative Perceived Usefulness | | | |
|---|---|---|---|---|---|---|---|---|---|
| Configuration | 1 | 2 | 3 | 4 | 5 | 6 | 7 | 8 | 9 |
| **Activities of Daily Living** | | | | | | | | | |
| Travelling | ⊗ | | ⊗ | ⊗ | • | ● | ● | ⊗ | ● |
| Dressing | ⊗ | ● | • | • | ⊗ | ⊗ | • | • | • |
| Feeding/Bathing/Toileting | ⊗ | ⊗ | | | ⊗ | ⊗ | ⊗ | ● | • |
| **Intelligent Assistive Devices Usage** | | | | | | | | | |
| Access to Paid Apps | | ⊗ | ⊗ | ● | ● | ⊗ | ⊗ | ⊗ | ⊗ |
| **Demographics** | | | | | | | | | |
| Education | | ⊗ | ● | ⊗ | • | | ● | ⊗ | ● |
| Disability Grants | • | ● | • | • | ⊗ | ● | ⊗ | • | • |
| Consistency | 0.773 | 0.75 | 0.8 | 0.835 | 1 | 0.619 | 0.835 | 0.668 | 0.67 |
| Raw Coverage | 0.37 | 0.076 | 0.1 | 0.084 | 0.05 | 0.203 | 0.078 | 0.125 | 0.063 |
| Unique Coverage | 0.37 | 0.076 | 0.1 | 0.084 | 0.05 | 0.203 | 0.078 | 0.125 | 0.063 |
| Overall Solution Consistency | 0.795 | | | | | 0.667 | | | |
| Overall Solution Coverage | 0.681 | | | | | 0.469 | | | |

*Note: Black circles (●) indicate the presence of a condition; crossed-out circles (⊗) indicate its absence. Blank fields indicate "do not care" conditions. Larger circles indicate core conditions; smaller circles indicate peripheral ones (Fiss, 2011).*

Therefore, the ability of this group to independently manage daily life is low. Yet, in contrast to the previous group (4-5), the absence of access to paid apps was found to be a core condition across all four configurations. Meanwhile, demographic factors, such as education and disability grants, did not yield identifiable patterns.

## 5. Discussion

### 5.1. Implications

Based on the configurations from the QCA-based analysis, and taking the lens of SMD theory, we find that the individual support required for VIPs differs (Alma et al., 2012). Key differentiators are the degree of independence of VIPs, measured by the number and types of ADLs they need assistance with, and the ability to access IATs. Whereas the grant itself does not provide enough support to positively impact IAT usefulness, it is instead the local community and social network of VIPs (Kef et al., 2000) that determine whether IATs can enable VIPs to have greater quality of life by effectively assisting them in their ADLs. We find three main implications for future research and policymakers on the intersection of VIPs and IATs. We discuss these findings by providing relevant quotes (note ID1-37 are survey-based; ID38-61 are interview) and reference related research.

**Financial Aid.** Our analysis reveals that disability grants do not have a significant impact on any of the three groups of configurations we identified. We argue that the disability grant, for those who receive it, merely covers the necessities such as groceries and e-hailing costs, and therefore, does not have a measurable impact on IATs. Rather, those who can access loans or funds from relatives ("*I use my brother's laptop, currently I can't afford it. So I just I use shortcuts with Jaws*"; ID55) or their local communities ("*I am fortunate that I got all my devices from blind organizations*"; ID46) can access IAT devices. In line with existing research, VIPs typically rely on individuals providing primary support that help them to perform their daily life tasks (Silva-Smith et al., 2007). The government therefore needs to provide more financial support to have a measurable effect on the perceived usefulness of IAT devices and ultimately support their independence: "*A braille note taker would change the life of a blind person unfortunately the one I think of is like R20 000 […] they really give us independence* (ID38)."

**Personalized Support.** The analysis further shows that there is no panacea to address the situation of VIPs. In line with Alma et al. (2012), who find that individual factors predict quality of life, personalized support tailored to the individual's needs are required. Consider the following statement: "*I think it must be suitable for my needs, like, Talkback does not read Afrikaans if I receive a text. I got used to Afrikaans on how it sounds. As I was not born blind, I know how to spell, I would listen to each character even though it is not reading Afrikaans* (ID39)." In this case, two factors predict the usefulness of IATs, (1) the support for the spoken language; and (2) whether the VIP was born blind or not. Despite factor 1 not being given, the IAT is still useful due to the mediating capability of factor 2. Many VIPs have mentioned the difficulty of transferring from one place to another. Again, we find that individual needs vastly depend on local specifics: "*Here in Eastern Cape, I don't feel accommodated as a blind person, the only place I saw is accommodative was Worcester in Western Cape* (ID42)."

**Social Network.** Lastly, we find that the knowledge about IATs helpful to VIPs for their personal circumstances largely depends on the availability of social ties within and support from local communities. This is consistent with previous findings, who found that social networks of VIPs are more impactful on well-being than individual characteristics (Kef et al., 2000). Typically, knowledge about the usefulness about IATs spreads by word of mouth: "*As blind people, we share information, there is a guy who shares with me these new devices, so I would just explore them* (ID52)." Therefore, those who live in an environment without an established community might lack information about IATs useful to them and hence might rate the perceived usefulness of IATs lower.

## 5.2. Limitations and Future Research

There are a few limitations to this study. Firstly, given that the focus of this study was to investigate accessibility of IATs, this study only investigated perceived usefulness as outcome variable. However, this only constitutes one possible antecedent of an individual's intention to use IATs from the perspective of the Technology Acceptance Model (TAM) (Davis, 1989). Future research is encouraged to explore the effect of ease of use across different IATs (Table 1).

Secondly, this study only measured five basic ADLs in Table 3 and excluded the instrumental activities of daily living (IADLs). It would be valuable for future research to measure the full spectrum of functional capabilities using both ADLs and IADLs.

Thirdly, semi-structured interviews were conducted via telephone. Therefore, we had to exclude deaf-blind people (DBP). We suggest future research to address how DBPs utilize IATs on their ADLs.

**Acknowledgements**. This research was funded in part by South Africa's National Research Foundation (NRF) under the reference MND210607608912 and the Luxembourg National Research Fund (FNR) and PayPal, PEARL grant reference 13342933/Gilbert Fridgen. For the purpose of open access, and in fulfillment of the obligations arising from the grant agreement, the author has applied a Creative Commons Attribution 4.0 International (CC BY 4.0) license to any Author Accepted Manuscript version arising from this submission.## 6. References

Addo, E. K., Akuffo, K. O., Sewpaul, R., Dukhi, N., Agyei-Manu, E., Asare, A. K., Kumah, D. B., Awuni, M., & Reddy, P. (2021). Prevalence and associated factors of vision loss in the South African National Health and Nutrition Examination Survey (SANHANES-1). BMC Ophthalmology, 21(1), 1.

Alma, M. A., Van der Mei, S. F., Groothoff, J. W., & Suurmeijer, T. P. (2012). Determinants of social participation of visually impaired older adults. Quality of Life Research, 21, 87-97.

Aminparvin, H., Henrichs, L., Auger, C., Dumassais, S., Renaud, J., & Wittich, W. (2025). Follow-up in low vision rehabilitation for users of assistive technology: a scoping review. Disability and Rehabilitation: Assistive Technology, 20(4), 721-732.

Balakrishnan, M. (2022). Computing and assistive technology solutions for the visually impaired. Communications of the ACM, 65(11), 20–22.

Cao, Shiya; Loiacono, Eleanor; and Annabi, Hala, "A Design Framework for Information Systems in the Workplace Accommodation Process from a Social Model Perspective: A Research Plan" (2020). AMCIS 2020 Proceedings. 5.

Datt, S., Senapathi, M., & Mirza, F. (2017). IO Vision–an integrated system to support the visually impaired. Australasian Conference on Information Systems.

Davis, F. D. (1989). Perceived usefulness, perceived ease of use, and user acceptance of information technology. MIS quarterly, 319-340.

Edemekong, P.F., Bomgaars, D.L., Sukumaran, S., et al. (2023) Activities of Daily Living. StatPearls, Treasure Island.

Fiss, P. C. (2011). Building Better Causal Theories: A Fuzzy Set Approach to Typologies in Organization Research, Academy of Management Journal 54(2), 393–420.

Fota, A., & Schramm-Klein, H. (2024). The impact of digital voice assistants on the everyday life of physically disabled users: Barriers, drivers and potential for improvement. Proceedings of the 57th Hawaii International Conference on System Sciences.

Gao, H., Ng, E., Deng, B., & Chau, M. (2024). Are real-time volunteer apps really helping visually impaired people? A social justice perspective. Information & Management, 61(6), 104007.

Haegele, J. A., & Hodge, S. (2016). Disability Discourse: Overview and Critiques of the Medical and Social Models. Quest, 68(2), 193–206.

Haesebrouck, T. (2016). The added value of multi-value qualitative comparative analysis. In Forum Qualitative Sozialforschung/Forum: Qualitative Social Research 17(1).

Hubner, S., Blaskewicz Boron, J., & Fruhling, A. (2022). Use of assistive and interactive technology and relation

to quality of life in aging adults. Hawaii International Conference on System Sciences, 1664.

Joshi, P., & Pappageorge, J. (2023). Reimagining disability: A call to action. Developmental Disabilities Network Journal, 3(2), 7.

Katz, S. (1983). Assessing self-maintenance: activities of daily living, mobility, and instrumental activities of daily living. Journal of the American Geriatrics Society, 31(12), 721-727.

Kef, S., Hox, J. J., & Habekothe, H. T. (2000). Social networks of visually impaired and blind adolescents. Structure and effect on well-being. *Social Networks, 22*(1), 73-91.

Khan, A., & Khusro, S. (2021). An insight into smartphone-based assistive solutions for visually impaired and blind people: issues, challenges and opportunities. Universal Access in the Information Society, 20(2), 265-298.

Kim, K., Ren, X., Choi, S., & Tan, H. Z. (2016). Assisting people with visual impairments in aiming at a target on a large wall-mounted display. International Journal of Human-Computer Studies, 86, 109–120.

Kugler, L. (2020). Technologies for the visually impaired. Communications of the ACM, 63(12), 18–20.

Lo, C.-L., Yang, X.-R., & Tseng, H.-T. (2024). Utilizing Assistive and Interactive Technologies: Impacts on Quality of Life among Aging Adults. Pacific Asia Conference on Information Systems.

Mattke, J., Maier, C., Weitzel, T., Gerow, J., & Thatcher, J. (2022). Qualitative Comparative Analysis (QCA) In Information Systems Research: Status Quo, Guidelines, and Future Directions. Communications of the Association for Information Systems, 50(1).

McGrath, C., Galos, Y., Bassey, E., & Chung, B. (2025). The influence of assistive technologies on experiences of risk among older adults with age-related vision loss (ARVL). Disability and Rehabilitation: Assistive Technology, 20(1), 118-126.

Muñoz, E., Serrano, M., Vivó, M., Marqués, A., Ferreras, A., & Solaz, J. (2016). SIMON: Assisted Mobility for Older and Impaired Users. Transportation Research Procedia, 14, 4420–4429.

Olkin, R. (2002). Could you hold the door for me? Including disability in diversity. Cultural Diversity and Ethnic Minority Psychology, 8, 130-137.

Oliver, M. (1983). Social Work with Disabled People. Basingstoke: Macmillan.

Oliver, M. (2013) The social model of disability: thirty years on, Disability & Society, 28:7, 1024-1026.

Pappas, I. O., & Woodside, A. G. (2021). Fuzzy-set Qualitative Comparative Analysis (fsQCA): Guidelines for research practice in Information Systems and marketing. International Journal of Information Management, 58, 102310.

Pizarro-Pennarolli, C., Sánchez-Rojas, C., Torres-Castro, R., Vera-Uribe, R., Sanchez-Ramirez, D. C., Vasconcello-Castillo, L., & Rivera-Lillo, G. (2021). Assessment of activities of daily living in patients post COVID-19: a systematic review. PeerJ, 9, e11026.

Ragin, C. C. (2009). Redesigning Social Inquiry: Fuzzy Sets and Beyond. University of Chicago Press.

Ragin, C. C., & Davey, S. (2022). Fuzzy-Set/Qualitative Comparative Analysis (Version 4.0). https://sites.socsci.uci.edu/~cragin/fsQCA/citing.shtm

Rodriguez-Sancheza, M. C., & Martinez-Romo, J. (2017). GAWA – Manager for accessibility Wayfinding apps. International Journal of Information Management, 37(3), 505–519.

Sanderson, L., Campbell-Meier, J., Goulding, A., & Sylvester, A. (2024). "It makes my life worthwhile": Digital inclusion, digital equity, and digital justice for people with disabilities. In ACIS 2024 Proceedings, 56.

Senjam, S. S. (2021). Smartphones as assistive technology for visual impairment. Eye, 35(7), 2078–2080.

Shethia, S., Waizenegger, L., & Techatassanasoontorn, A. A. (2023). " Goodnight Alexa"–Theorising interactions between people with visual impairments and digital voice assistants. International Conference on Information Systems, 16.

Silva-Smith, A. L., Theune, T. W., & Spaid, P. E. (2007). Primary support persons for individuals who are visually impaired: Who they are and the support they provide. Journal of visual impairment & blindness, 101(2), 113-118.

Theodorou, P., & Meliones, A. (2019). Developing apps for people with sensory disabilities, and implications for technology acceptance models. Global Journal of Information Technology: Emerging Technologies, 9(2), 33-40.

Tsibolane, P., & Nombakuse, R. (2024). Digital Empowerment Through Intelligent Assistive Technologies for Visually Impaired People. SIGSI 2024.

Vieira, A. D., Leite, H., & Volochtchuk, A. V. L. (2022). The impact of voice assistant home devices on people with disabilities: a longitudinal study. Technological Forecasting and Social Change, 184, 121961.

Vimalan, N., Zimmer, M. P., & Drews, P. (2024). Fostering inclusivity – Towards principles for inclusive information systems design. In Wirtschaftsinformatik 2024 Proceedings, 85.

World Health Organisation. (2023). Increasing eye care interventions to address vision impairment. https://www.who.int/publications/m/item/increasing-eye-care-interventions-to-address-vision-impairment

World Health Organisation. (2024). Eye Health. https://www.afro.who.int/health-topics/eye-health

Vink, M. P., & Van Vliet, O. (2009). Not quite crisp, not yet fuzzy? Assessing the potentials and pitfalls of multi-value QCA. Field Methods, 21(3), 265-289.

Xie, I., Babu, R., Lee, T. H., Davey Castillo, M., You, S., & Hanlon, A. M. (2020). Enhancing usability of digital libraries: Designing help features to support blind and visually impaired users. Information Processing and Management, 57 (102110).

Zhu, H., Samtani, S., Brown, R., & Chen, H. (2020). A deep learning approach for recognizing activity of daily living (ADL) for senior care: Exploiting interaction dependency and temporal patterns. MIS Quarterly, 45(2).